# Time delay in electron collision with a spherical target as a function of the scattering angle

M. Ya. Amusia[1, 2], A. S. Baltenkov[3]
and
I. Woiciechowski[4, a]

[1] *Racah Institute of Physics, the Hebrew University, 91904, Jerusalem, Israel*
[2] *Ioffe Physical-Technical Institute, 194021, St. Petersburg, Russian Federation*
[3] *Arifov Institute of Ion-Plasma and Laser Technologies,
100125, Tashkent, Uzbekistan*
[4] *Alderson Broaddus University, 101 College Hill Drive, WV 26416, Philippi, USA*

**Abstract**

We have studied the *angular time delay* in slow-electron elastic scattering by spherical targets as well as the *average time delay* of electrons in this process. It is demonstrated how the *angular time delay* is connected to the Eisenbud-Wigner-Smith (EWS) time delay. The specific features of both angular and energy dependences of these time delays are discussed in detail. The potentialities of the derived general formulas are illustrated by the numerical calculations of the time delays of slow electrons in the potential fields of both absolutely hard sphere and delta-shell potential well of the same radius. The studies conducted shed more light on the specific features of these time delays.

## 1. Introduction

In the first experiments, the purpose of which was to study the time delays of electrons in atomic photoeffect, electrons with the wave vector **k** emitted along the polarization vector **e** of the absorbed photon are recorded [1-3]. With this experimental technique, the delay times of the electrons escaping at an arbitrary angle to the vector **e** were unknown. Now, investigation of time delays as a function of the emission angle $\vartheta$ have become available [4-7], and the corresponding calculations have been able to reproduce this dependence for different atoms [8-13]. The electron delay time is a function depending on both the photoelectron emission angle $\vartheta$ with respect to the radiation polarization vector **e** and the photoelectron energy $E$. In most calculations of the time, its dependence on the energy $E$ is analyzed at fixed values of the angle $\vartheta$, revealing the pronounced angle dependence for large emission angles.

The angular dependence of the time delay of the wave packet scattered (or emitted) by a spherical target was obtained by Froissard, Goldberger, and Watson in [14], where the following expression for the angular time delay of the packet scattered in the direction $\vartheta$ was derived:

$$\Delta t(k, \vartheta) = \hbar \frac{\partial}{\partial E} \arg f(k, \vartheta), \quad \vartheta \neq 0. \tag{1}$$

Here $f(k, \vartheta)$ denotes the amplitude of electron elastic scattering by a target [15]

$$f(k, \vartheta) = \frac{1}{2ik} \sum_l (2l+1)(e^{2i\delta_l} - 1) P_l(\cos \vartheta), \tag{2}$$

[a]*Corresponding author.* E-mail address: woiciechowskiia@ab.edu

$\delta_l(k)$ is the partial scattering phase shifts, $P_l(\cos\vartheta)$ are the Legendre polynomilas. According to (1) the forward scattering $\vartheta = 0$ must be excluded due to the interference effects between the forward scattered wave and the incident wave that give rise to the optical theorem [15].

The domain of applicability of the *angular time delay* $\Delta t(k,\vartheta)$ (1) is considerable broader than that of the Eisenbud-Wigner-Smith (EWS) *partial-wave time delay* [16-18]

$$\tau_l(k) = 2\hbar \frac{\partial \delta_l}{\partial E}. \tag{3}$$

In particular, Eq. (1) serves as the basis for describing the temporal picture of atomic photoionization processes [17-25]. Eq. (1) in this case needs not to be modified to exclude $\vartheta = 0$, since the problem of the interference with the unscattered wave does not exist in the case of photoionization. The scattering amplitude $f(k,\vartheta)$ for this process must be replaced in Eq. (1) by the photoionization amplitude $f^{ph}(\omega,\vartheta)$ where $\omega$ is the photon energy

$$\tau(\omega,\vartheta) = \hbar \frac{\partial}{\partial \omega} \arg f^{ph}(\omega,\vartheta). \tag{4}$$

The dipole selection rules in photoionization of *l*-states of atom A lead to emission into continuum of the pair of electronic spherical waves $Y_{l+1,m}(\mathbf{k})$ and $Y_{l-1,m}(\mathbf{k})$, propagating in the potential field of the atomic residue A$^+$ with the phase shifts $\delta_{l+1}(k)$ and $\delta_{l-1}(k)$, correspondingly, where $k$ is the linear photoelectron momentum. The function $f^{ph}(\omega,\vartheta)$, therefore, is a linear combination of these spherical functions, the coefficients of which are determined by the corresponding dipole matrix elements $D_{l\pm 1}(\omega)$. The energy derivative of the function (3) implicitly includes the derivatives of both phase shifts $\delta'_{l\pm 1}(k)$ and matrix elements $D'_{l\pm 1}$. The prime sign here and further denotes differentiation with respect to the electron kinetic energy $E$.

The time delay (4) at some electron emission angles $\vartheta$ was studied in the series of works on *photoionization* [19-24]. To the best of our knowledge, the angular dependence of the time delay in elastic *electron scattering* (1) received no attention so far. Our goal in this article is to close somewhat the gap in the area of investigation of the angular time delay in electron scattering (1) by spherical targets.

We will see further that when only one scattering phase is different from zero in the scattering amplitude (2), the angular time delay (1) does not depend on the scattering angle. Here we analyze the scattering amplitude $f(k,\vartheta)$ containing two Legendre polynomials only, i.e., we will consider model targets, in which, as in the case of the dipole photoelectric effect, only one pair of phase shifts is different from zero.

In the next Section 2, the angle dependence of the angular time delay $\Delta t(k,\vartheta)$ for some fixed electron momenta $k$ is investigated. In Section 3, the time delay is studied as a function of $k$ for some fixed polar angles $\vartheta$ of the scattering of an incident plane wave train. Finaly, the function $\Delta t(k,\vartheta)$ is averaged over the distance of the order of the de Broglie wavelength, and the *average angular time delay* $\langle \Delta t(k) \rangle$ is obtained in the Section 4.

## 2. Angular $\vartheta$-dependence of the function $\Delta t(k,\vartheta)$

The argument of the amplitude $f(k,\vartheta)$ is determined by the ratio of the imaginary part of function (2) $\operatorname{Im} f(k,\vartheta)$ to its real part $\operatorname{Re} f(k,\vartheta)$

$$\arg f(k,\vartheta) = \arctan \frac{\operatorname{Im} f(k,\vartheta)}{\operatorname{Re} f(k,\vartheta)}, \tag{5}$$

whereas the angular time delay (1) is described by the general expression

$$\Delta t(k,\vartheta) = \frac{d}{dE} \arg f(k,\vartheta) = \frac{(\operatorname{Im} f)'(\operatorname{Re} f) - (\operatorname{Re} f)'(\operatorname{Im} f)}{|f|^2}. \tag{6}$$

Here and everywhere below, we use the atomic system of units.

Let us first consider the case when all the phase shifts in (2), with the exception of $\delta_l(k)$, are equal to zero. In this case

$$f(k,\vartheta) = \frac{1}{2k}(2l+1)P_l(\cos\vartheta)\sin 2\delta_l + i\frac{1}{k}(2l+1)P_l(\cos\vartheta)\sin^2 \delta_l,$$

$$\arg f(k,\vartheta) = \arctan[\tan \delta_l], \tag{7}$$

$$\Delta t(k,\vartheta)_l = \frac{d\delta_l}{dE} \equiv \delta_l'.$$

It is seen that the angular time delay does not depend on the scattering angle $\vartheta$, and it is equal to half of the EWS-partial time delay (3).

Suppose that only two scattering phases $\delta_0(k)$ and $\delta_1(k)$ are nonzero. In this case, the scattering amplitude and its argument are represented as

$$f(k,\vartheta) = \frac{1}{2k}(P_0 \sin 2\delta_0 + 3P_1 \sin 2\delta_1) + i\frac{1}{k}(P_0 \sin^2 \delta_0 + 3P_1 \sin^2 \delta_1)$$

$$\arg f(k,\vartheta) = \arctan \frac{2(P_0 \sin^2 \delta_0 + 3P_1 \sin^2 \delta_1)}{(P_0 \sin 2\delta_0 + 3P_1 \sin 2\delta_1)}, \quad P_{0,1} \equiv P_{0,1}(\cos\vartheta). \tag{8}$$

Differentiating the argument of the scattering amplitude (8), we obtain the expression for the time delay

$$\Delta t(k,\vartheta)_{01} =$$

$$\frac{P_0[P_0 \sin^2 \delta_0 + 3P_1 \sin(2\delta_0 - \delta_1)\sin \delta_1]\delta_0' + 3P_1[3P_1 \sin^2 \delta_1 + P_0 \sin(2\delta_1 - \delta_0)\sin \delta_0]\delta_1'}{P_0^2 \sin^2 \delta_0 + 6P_0 P_1 \sin \delta_0 \sin \delta_1 \cos(\delta_0 - \delta_1) + 9P_1^2 \sin^2 \delta_1} \tag{9}$$

as a function of both scattering angle $\vartheta$ and electron momentum $k = \sqrt{2E}$.

Repeating the calculations similar to those in formulae (8), we obtain the expression for the time delay in the case of nonzero phases $\delta_0(k)$ and $\delta_2(k)$

$$\Delta t(k,\vartheta)_{02} =$$

$$\frac{P_0[P_0 \sin^2 \delta_0 + 5P_2 \sin(2\delta_0 - \delta_2)\sin \delta_2]\delta_0' + 5P_2[5P_2 \sin^2 \delta_2 + P_0 \sin(2\delta_2 - \delta_0)\sin \delta_0]\delta_2'}{P_0^2 \sin^2 \delta_0 + 10 P_0 P_2 \sin \delta_0 \sin \delta_2 \cos(\delta_0 - \delta_2) + 25 P_2^2 \sin^2 \delta_2}$$
(10)

It is easy to demonstrate that when only two scattering phases $\delta_l(k)$ and $\delta_{l'}(k)$ are nonzero in the electron scattering amplitude (2), the angular delay time (5) is determined by the following combination of the Legendre polynomials $P_l(\cos\vartheta)$ and $P_{l'}(\cos\vartheta)$:

$$\Delta t(k,\vartheta)_{ll'} = \frac{\sum_{i=l,l'}\sum_{j=l,l'}[(2i+1)(2j+1)P_i P_j \sin(2\delta_i - \delta_j)\sin \delta_j]\delta_i'}{\sum_{i=l,l'}\sum_{j=l,l'}(2i+1)(2j+1)P_i P_j \sin \delta_i \sin \delta_j \cos(\delta_i - \delta_j)}.$$
(11)

Explicit expressions for the time delays for selected nonzero scattering phase pairs (11) are given in [26], where the results of the calculations of the $\vartheta$- and $E$-dependences of the corresponding angular time delays are also given.

We use further the hard sphere- and delta-shell potentials as potential functions for the model targets. For these potentials, the analytical expressions for the scattering phases are known. When an electron is scattered by the model target in the form of an ideally repulsive solid sphere of radius $R$, the phase shifts of the electron are determined by the formula [27]

$$\tan \delta_l(k) = \frac{j_l(kR)}{n_l(kR)},$$
(12)

where $j_l(kR)$ and $n_l(kR)$ are the spherical Bessel functions.

The scattering phase shifts of an electron for another model target taken in the form of an attractive delta-shell (delta-shell potential well [28]) are determined by the expression (see Eq. (10) in [29])

$$\tan \delta_l = \left[\frac{xj_l^2(x)}{xj_l(x)n_l(x) - 1/R\Delta L}\right],$$
(13)

where the variable $x = kR$. The parameter $\Delta L$ in (13) is the jump of the logarithmic derivative of the electron wave functions at the point $r = R$ where the delta-shell potential $U(r) = -U_0 \delta(r - R)$ is infinitely negative. In the numerical calculations of phase shifts (12) and (13), the radii $R$ and the parameter $\Delta L$ have the same values as those used in our article [29], where the EWS time delay of slow electrons scattered by a $C_{60}$ cage was calculated.

Figure 1 shows the results of the calculation by formula (9) of the angular time delay $\Delta t(k,\vartheta)_{01}$ as a function of the scattering angle $\vartheta$ for some fixed electron momenta $k$. The left panel corresponds to the scattering on the solid sphere, the right one does to the delta-shell target. The angular time delays in these figures are given in atomic units. The atomic unit of time is equal to 24.2 attoseconds. Despite the different scales of the graphs on both panels, they show qualitatively similar behavior.

The only exceptions are for the curves at $k = 0.68$. In the case of hard sphere, the graph of the angular dependence is almost a straight line, passing from a positive to a negative half-plane at the angle of about 60°. Whereas on the right panel, this curve almost coincides with the x-axis. According to both panels, at low electron energies ($k = 0.17$ and $0.34$), the time delay of the scattering packet is negative at all the scattering angles. The rest of the curves are alternating for both targets. At the momenta $k = 0.51$ and $k = 1.0$, the time delays on the right panel reach a maximum (~ 298 atomic units (au) at $\vartheta = 95°$ in the first case and ~ 140 au at the same angle in the second one). The appearance of these sharp peaks in the curves in Figure1 is due to the almost vanishing of the denominator in the expression (9). The curves at $k = 0.85$ and $k = 1.0$ on the left panel cross the x-axis into the positive half-plane in the region of 90°, forming a peak with a height of about 30 atomic units.

Figure 2 depicts the curves corresponding to the pair of polynomials $P_0(\cos\vartheta)$ and $P_2(\cos\vartheta)$. We see here the results of the calculation with formula (10) of the angular time delay $\Delta t(k,\vartheta)_{02}$ as a function of the scattering angle $\vartheta$. Since the sum of the orbital moments (indices of the Legendre polynomials) is an even number, the curves $\Delta t(k,\vartheta)_{02}$ in Figure 2 are symmetric relative to the angle $\vartheta = 90°$. The curves on the left panel, except for the curve at $k = 0.34$, lie entirely in the lower half-plane. The situation is quite different when the packet scatters by the delta-shell target. The behavior of the curve at $k = 1.0$ on the right panel is particularly interesting. This curve lies entirely in the positive half-plane, which allows it to be depicted in polar coordinates (see the inset in the right panel). The 3D-picture of the function $\Delta t(k,\vartheta)_{02}$ is a figure of rotation of this curve around the polar axis $z$, along which the incident plane wave train hits the target. The "wings of the star" shown there correspond to the polar scattering angles $\vartheta = 57°$ and $123°$. The qualitative similarity of the curves on both panels of Figure 2 is obvious.

Let us pay attention to the similarity of the curves in Figure 2 and the angular spectra in Figure 1 (a) and (b) of article [10] devoted to the study of angular resolved time delays in photoemission from different atomic subshells of noble gases. Photoelectron spectra are similar to the scattering spectra being symmetric relative to the angle $\vartheta = 90°$. Qualitative behavior of the scattering spectrum on the delta-shell at $k = 0.51$ in Figure 2 and the photoelectron spectrum in panel (a) of Figure 1 is similar. The same can be stated about the curves at $k = 0.85$ in Figure 2 and the curves in panel (b) of Figure 1 in [10].

Summarizing, we note that according to Figures 1 and 2, the angular $\vartheta$-dependences of the function $\Delta t(k,\vartheta)$ are represented by nontrivial rapidly oscillating curves lying at low electron energies mainly in the negative half-plane. The situation changes with increasing the electron energy.

## 3. $k$-dependence of function $\Delta t(k,\vartheta)$

In this section, we investigate the angular time delay $\Delta t(k,\vartheta)$ as a function of the electron energy $E$ for some fixed values of polar angles $\vartheta$. The calculation results by formulas (9) and (10) are shown in Figure 3. All curves in this Figure tend to infinity at small electron momenta. The reason for this is that the scattering phase shift in short-range potentials should follow the Wigner threshold law $\delta_l(E) \propto E^{l+1/2}$ [30]. In the case of $s$-phase shift, we have $\delta_0(E \to 0) \propto \pi - E^{1/2}$. So, the time delay $\Delta t(k,\vartheta)_{01}$ and $\Delta t(k,\vartheta)_{02}$, that contain the derivative of $s$-phase shift, for $k \to 0$ tends to infinity:

$\delta'_0(E \to 0) \propto -E^{-1/2}$. For the orbital moments $l > 0$ the derivative of phase shifts should be vanishing at the threshold as $\delta'_l(E \to 0) \propto E^{l-1/2}$.

The left column of the figures corresponds to the electron scattering by the hard sphere potential. The figures in the right column correspond to scattering by the delta-shell potential. In the upper right panel of Figure 3, the curves practically coincide with each other at small scattering angles $\vartheta$, up to the angle of 45°. The graphs corresponding to the angles of 135° and 180° have alternating signs, and they are characterized by the peaks in both positive and negative half-planes of the coordinate system. We see the qualitatively similar picture in lower panel of this colomn, where the curves for $\Delta t(k,\vartheta)_{02}$ are presented. The presence of the derivative of the *s*-phase shift in formula (10) also leads this function to infinity at small electron energies. The curves for angles 30° and 180° almost coincide in this figure. The curve at $\vartheta = 90°$ is characterized by the maximum negative amplitude of oscillations. In the lower left panel of Figure 3, we observe strong resonance behavior of all curves, except for the one at $\vartheta = 90°$, and energy $E \approx 0.4$ atomic units.

In the second and third sections, we limited ourselves to the specific examples of two nonzero phases in the expansion of the wave function of a scattered electron (2) into partial waves. We see that even this simplest example leads to a very difficult for interpretation, rapidly oscillating dependence of the time delays upon the energy $E$ and scattering angle $\vartheta$. An increase in the number of included essential scattering phases significantly affects the picture of the angular time delay, making it so rapidly oscillating that its averaging over the energy of incident electrons and the scattering angle becomes inevitable in order to make the angular time delay $\Delta t(k,\vartheta)$ observable in an experiment.

### 4. Average time delay of the scattering process.

The average angular time delay $\langle \Delta t(k) \rangle$ is obtained from (1) by averaging over the energy spectrum of the incident wave packet, as well as over the directions weighted by the differential cross section $|f(k,\vartheta)|^2$. This averaging is reduced to calculation of the integral of the product $|f(k,\vartheta)|^2 \Delta t(k,\vartheta)$ over all angles of electron scattering by the target and division of the obtained result by the total cross section of elastic electron scattering $\sigma_{tot}(k)$. The calculation of the integral is complicated by the fact that, according to (1), the function $\Delta t(k,\vartheta)$ is not defined at $\vartheta = 0$. It was shown in [31] that the contribution to the integral from forward scattering of an electron is determined by the real part of the scattering amplitude at zero angle. As a result of such averaging, Nussenzweig [31-33] obtained the expression

$$\langle \Delta t(k) \rangle = \frac{1}{\sigma_{tot}(k)} \left( \int |f(k,\vartheta)|^2 \Delta t(k,\vartheta) d\Omega + \frac{2\pi}{k^2} \frac{d}{dE}[k \operatorname{Re} f(k,0)] \right) =$$
$$\frac{\pi}{\sigma_{tot}(k)k^2} \sum_l (2l+1) 2\delta'_l = \frac{\pi}{\sigma_{tot}(k)k^2} \sum_l (2l+1) \tau_l(k)$$
(14)

The second term on the left-hand side of the equation (14) eliminates the contribution of the forward scattering into the average angular time delay. Thus, the average time delay for the plane wave train $\langle \Delta t(k) \rangle$ is a linear combination of the EWS time delays $\tau_l(k)$ (3).

The results of the calculation of the function $\langle \Delta t(k) \rangle$ (14) in the case of electrons scattered by the hard sphere target are shown in Figure 4. Figure 4 also shows the dependencies calculated under the assumption that the statistical weight of $\tau_l(k)$ in the sum (14) is not equal to $\pi(2l+1)/\sigma_{tot}k^2$, but it is the ratio of the electron elastic scattering partial cross section $\sigma_l(k)$ to the total cross section $\sigma_l(k)/\sigma_{tot}(k)$. About this assumption see, for example, Eq. (10) in [10] or Eq. (8) in [29]. The deep peak of the curve corresponding to the combination of the Legendre polynomials $P_0$ and $P_2$ is due to the the resonant behavior of curves at $E \sim 0.4$ au in Fig. 3.

## 5. Concluding remarks

Using the instructive soluble example of electron scattering by the hard sphere potential and delta-shell potential well, we for the first time explicitly obtained the angular time-delay $\Delta t(k, \vartheta)$ in terms of the scattering phase shifts $\delta_l(k)$ and their energy derivatives $\delta'_l(k)$. We demonstrate the complexity of $\Delta t(k, \vartheta)$ as a function of the incoming electron energy $E$ and the scattering angle $\vartheta$. We see that $\Delta t(k, \vartheta)$ and the function $\langle \Delta t(k) \rangle$, even averaged over proper intervals of $E$ and $\vartheta$, are more sensitive to the scattering phases than the absolute cross section $\sigma_{tot}(k)$ and even the differential in angle scattering cross section that is proportional to $|f(k, \vartheta)|^2$. This is because the time delay functions depend not only on the cross-section phases, but upon their energy derivatives. This makes theoretical and experimental investigation of time delays a promising direction of research in the area of atomic scattering.

**Acknowledgment**
A. S. B. is grateful for the support to the Uzbek Foundation Award OT-2-46.

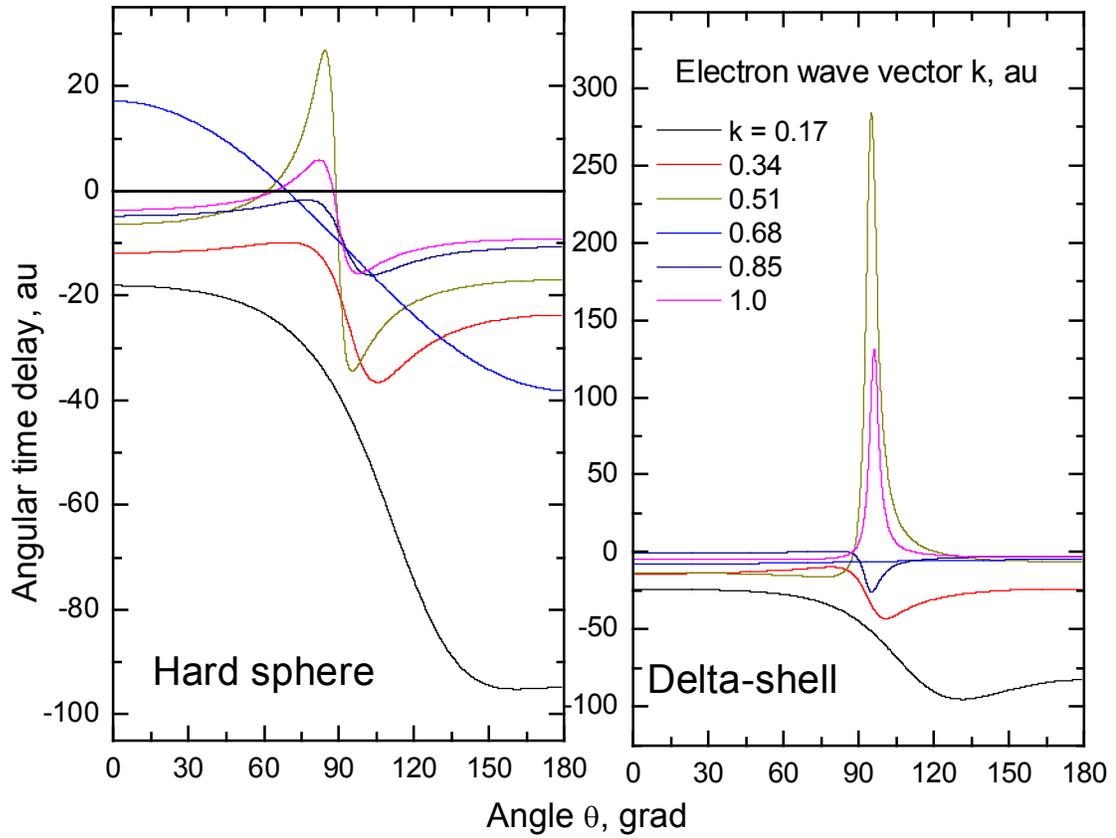

**Figure 1**. Angular time delay $\Delta t(k,\vartheta)_{01}$ (9) as a function of the polar angle $\vartheta$ for fixed electron wave vectors $k$. The functions $P_0(\cos\vartheta)$ and $P_1(\cos\vartheta)$ used in (9) is the pair of Legendre polynomials in the amplitude of electron elastic scattering $f(k,\vartheta)$ (2).

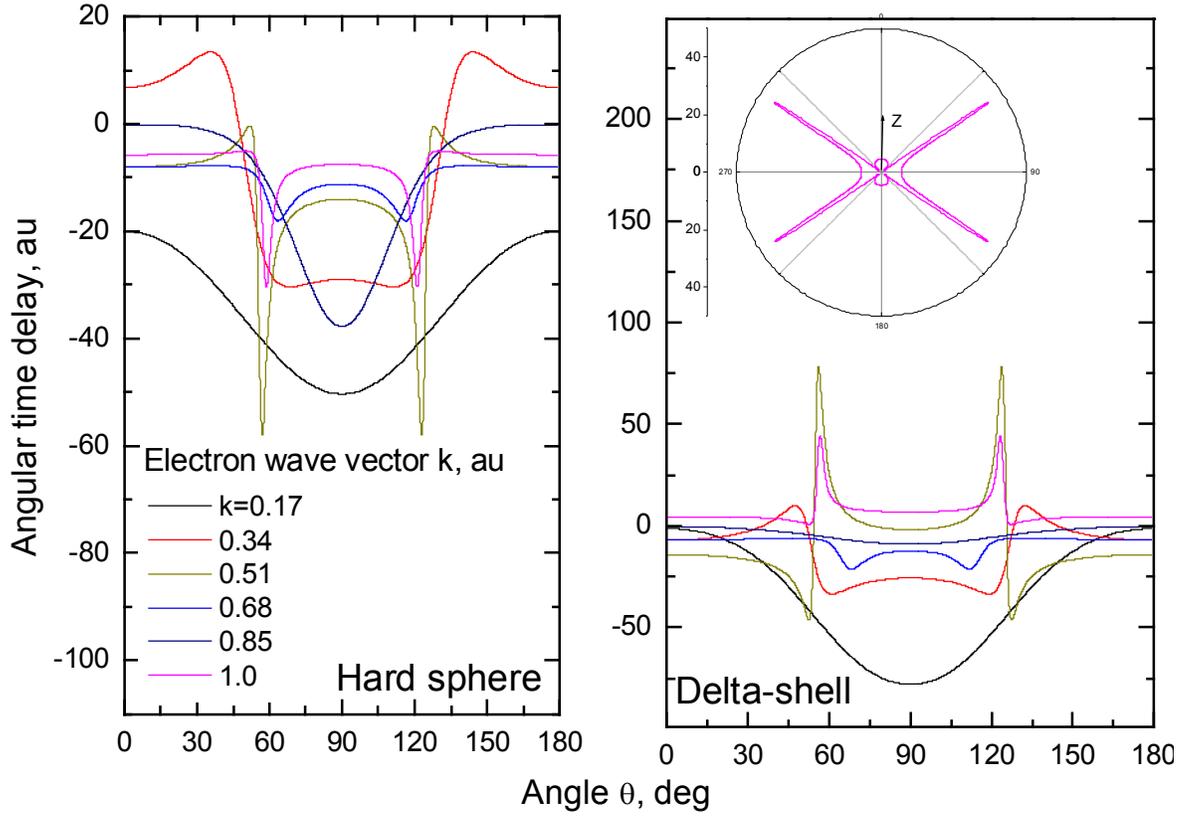

**Figure 2**. Angular time delay $\Delta t(k,\theta)_{02}$ (10) as a function of the polar angle $\vartheta$ for fixed electron wave vectors $k$. The functions $P_0(\cos\vartheta)$ and $P_2(\cos\vartheta)$ used in (10) is the pair of Legendre polynomials in the amplitude of electron elastic scattering $f(k,\vartheta)$ (2). The inset in the right panel is the plot at $k = 1.0$ in a polar coordinate system. The 3D-plot of the function $\Delta t(k,\vartheta)_{02}$ is a figure of rotation of this curve around the polar axis $z$, along which the incident plane wave train hits the target.

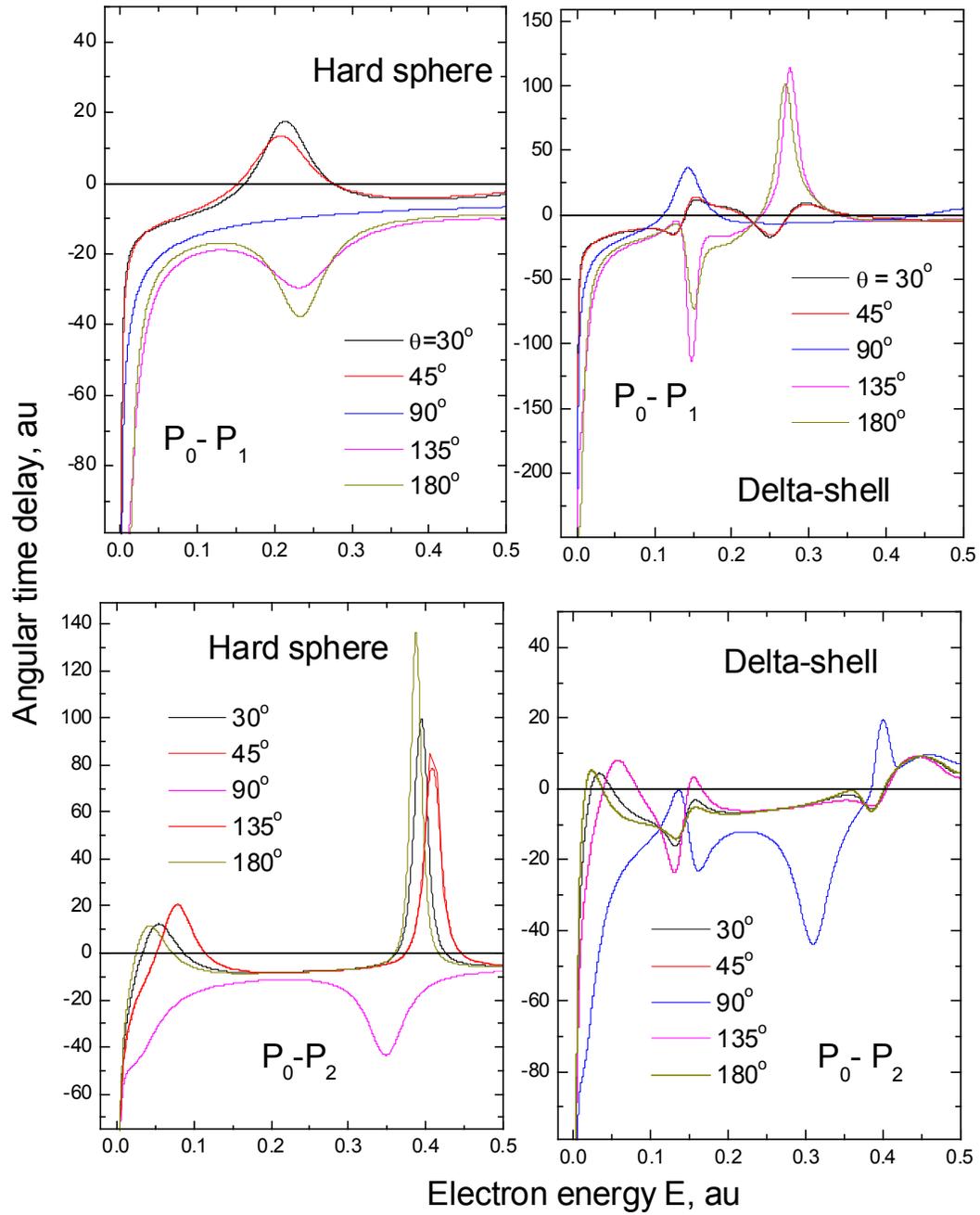

**Figure 3**. The angular time delay $\Delta t(k,\vartheta)$ as a function of the electron energy $E$ for some fixed values of the polar angle $\vartheta$. $P_0(\cos\vartheta)$–$P_1(\cos\vartheta)$ is the pair of Legendre polynomials in the upper panels. $P_0(\cos\vartheta)$–$P_2(\cos\vartheta)$ are the polynomials used in both lower panels.

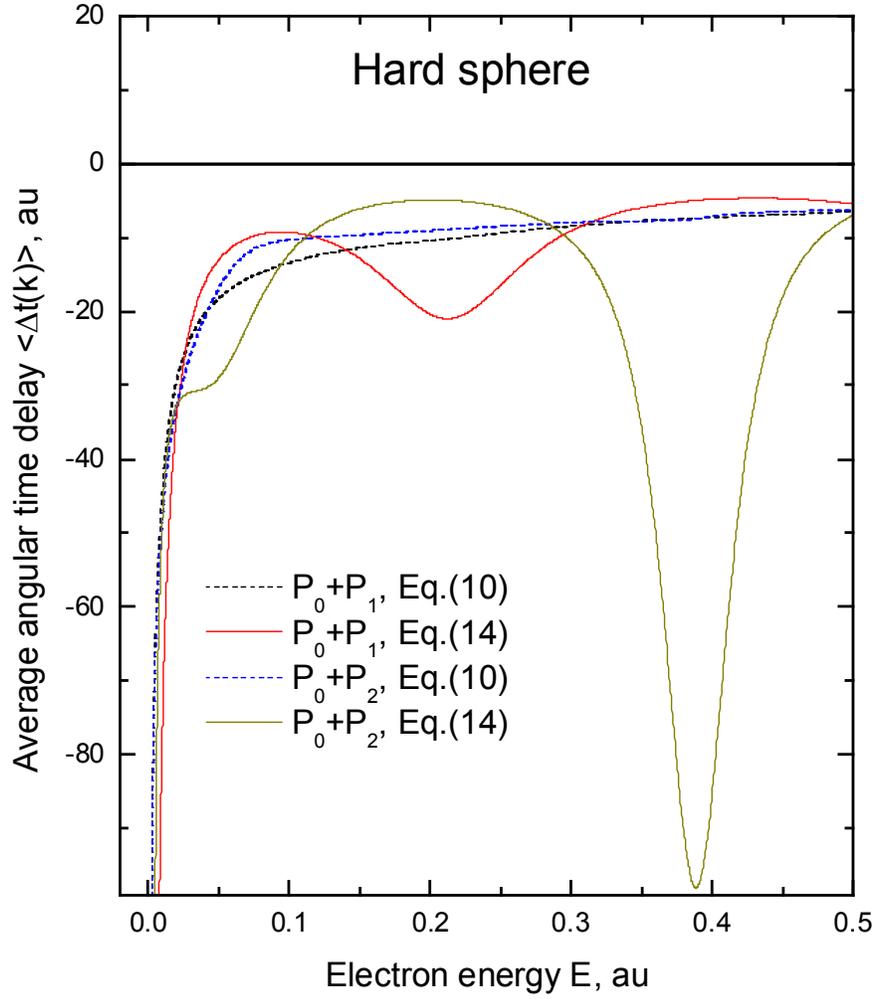

**Figure 4**. The function $\langle \Delta t(k) \rangle$ (14) in the case of electrons scattered by the hard sphere target. $P_0+P_1$ and $P_0+P_2$ are the pairs of Legendre polynomials $P_l(\cos \vartheta)$. Note that Equation (10) corresponds to Formula (10) in paper [10].